# Could Interaction with Social Robots Facilitate Joint Attention of Children with Autism Spectrum Disorder?

Wei Cao: First author; Centre for Optical and Electromagnetic Research, South China Academy of Advanced Optoelectronics, South China Normal University (SCNU); Science Building #5, Higher Education Mega Center Campus, South China Normal University, Guangzhou 510006, P. R. China.

Wenxu Song: First author; Centre for Optical and Electromagnetic Research, South China Academy of Advanced Optoelectronics, South China Normal University (SCNU); Science Building #5, Higher Education Mega Center Campus, South China Normal University, Guangzhou 510006, P. R. China.

Xinge Li: School of Psychology, South China Normal University (SCNU); School of Psychology, No 55, West Zhongshan Avenue, Tianhe District, South China Normal University, Guangzhou 510631, P. R. China.

Sixiao Zheng: School of Physics and Telecommunication Engineering, South China Normal University ,Guangzhou ,Guangdong 510006, P. R. China.

Ge Zhang: Caihongqiao children rehabilitation and service center of Panyu district, Panyu district, Guangzhou, P. R. China.

Yanting Wu: School of Psychology, South China Normal University (SCNU); School of Psychology, No 55, West Zhongshan Avenue, Tianhe District, South China Normal University, Guangzhou 510631, P. R. China.

Sailing He: Centre for Optical and Electromagnetic Research, South China Academy of Advanced Optoelectronics, South China Normal University (SCNU); Science Building #5, Higher Education Mega Center Campus, South China Normal University, Guangzhou 510006, P. R. China.

Huilin Zhu: Corresponding author; Children Developmental & Behavioral Center, Third Affiliated Hospital of Sun Yat-Sen University, Guangzhou 510630, P. R. China; Centre for Optical and Electromagnetic Research, South China Academy of Advanced Optoelectronics, South China Normal University (SCNU); Science Building #5, Higher Education Mega Center Campus, South China Normal University, Guangzhou 510006, P. R. China. (huilin.zhu@m.scnu.edu.cn)

Jiajia Chen: Corresponding author; Centre for Optical and Electromagnetic Research, South China Academy of Advanced Optoelectronics, South China Normal University (SCNU), Guangzhou 510006, P. R. China. KTH Royal Institute of Technology, Stockholm, Sweden. (jiajiac@kth.se)

**Abstract:** This research addressed whether interactions with social robots could facilitate joint attention of the autism spectrum disorder (ASD). Two conditions of initiators, namely "Human" vs. "Robot" were measured with 15 children with ASD and 15 age-matched typically developing (TD) children. Apart from fixation and gaze transition, a new longest common subsequence (LCS) approach was proposed to analyze eye-movement traces. Results revealed that children with ASD showed deficits of joint attention. Compared to the human agent, robot facilitate less fixations towards the targets, but it attracted more attention and allowed the children to show gaze transition and to follow joint attention logic. This results highlight both potential application of LCS analysis on eye-tracking studies and of social robot to intervention.

# Introduction

Autism spectrum disorder (ASD) is a neurodevelopmental disorder. Its overall prevalence is approximately 1/68 in North American (Centers for Disease Control and Prevention, 2016) and is similar in the developed region of China (Sun et al., 2015). Children diagnosed with ASD show persistent deficits in nonverbal communicative behaviors during social interaction, especially to maintain attention in eye contact, understand and follow the gaze direction (American Psychiatric Association, 2013). The treatment and intervention of ASD are usually very intensive, long-term and comprehensive undertaking, which involve the entire family and a team of professionals from multidiscipline. The ASD has become one of serious problems to the world health especially in the regions that the professionals are very limited.

Recently, more and more advanced technologies, such as humanoid social robot, have been involved in the diagnosis and intervention of the ASD (Diehl et al., 2012). Humanoid could provide similar social cues for interaction and communication as human, such as eye and head movements, gestures, and human-like voice. Therefore, whether humanoid could play an effective role in ASD intervention became an interesting question for researchers. On one hand, humanoid could offer many proper solutions to improve various abilities, including social behaviors (Damm et al., 2013), language (Kim et al., 2013), imitation (Cook et al., 2014), and it could reduce stereotyped behaviors (Shamsuddin et al., 2013). Researchers found that "ASD individuals had, toward robots, behaviors that typically developing (TD) individuals normally had toward human agents" (Pennisi et al., 2016). On the other hand, robots seemed to affect the interacting behaviors in a negative way. For example, Anzalone et al. (2014) used a commercial humanoid NAO to build an experimental platform, and their results reported that the school-aged children with ASD showed a decreased joint attention with NAO, but exhibited a similar performance as the age-matched TD groups when interacting with a human. Bekele et al. (2014) combined the robot NAO with a head tracker to build an adaptive robot intervention system, with which it has found that the ASD needed more prompts in the robot condition compared to the human condition in order to successfully find the target. Nevertheless, Warren et al. (2015) used NAO and an eye tracker to measure the joint attention behaviors of the ASD, and the results demonstrated that the prompts level decreased after four intervention sessions, which means the robot could improve ASD performance to respond to its gazes or points. Therefore, it remained an open question on whether and how humanoid could facilitate the social interaction skill of children with ASD.

Joint attention is one of the basis of social interactions, which "involves the use of procedures to co-ordinate attention between interactive social partners with respect to objects or events in order to share an awareness of the objects or events" (Mundy et al., 1986). Normally there are two kinds of joint attentions: Responsive joint attention (RJA) that refers to the ability to follow other's direction of gazes or points, and initial joint attention (IJA) that refers to the ability to share the attention or interest by using the direction of gazes or points (Mundy and Gomes, 1998; Mundy et al., 2009). Joint attention is a milestone of the early development of social communication, and plays a key role in the developmental deficits of the children with ASD, it is largely associated with the social communication and language acquisition, and hence has always been the target of the intervention of ASD (Charman, 2003).

A large quantity of researches have focused on the gaze following behaviors during joint attention and used eye-tracking technology to compare the difference of fixations and gaze transitions between children with ASD and TD. However, most of the previous studies did not yet find consistent results considering whether children with ASD showed the impairment in joint attention. Riby et al. (2013) used static pictures as the stimuli to distinguish the gaze following behavior of the school-aged high-functioning children with ASD. Their results revealed that compared to the non-verbal ability matched TD group, the ASD spent less time on the human face and the targets, but more on implausible objects, even with a cue. Falck-Ytter et al. (2012) used a Tobii 120 eye-tracker to measure the response of children with ASD (mean ages: 5.83±0.92 years), pervasive developmental disorder-not otherwise specified (PDD-NOS, 6.17±0.67) and TD controls (6.17±0.75) to the social scene, in which one actress gazed, pointed or both gazed and pointed at one out of three toys. The findings reported that the ASD group showed a significantly lower accuracy in following joint attention than the TD. Thorup et al. (2016) used a novel real life interaction experiment to measure the eye movement of infants (10.25±0.45 months) with ASD risk, infants with high-risk behaved worse when the actor only turned her eye. Those researches showed that there were significant impairments of joint attention skills in children with ASD. However, Billeci et al. (2016) used eye tracking technology to measure the RJA performance and IJA of the ASD (2~3 years) and age-matched controls. Results showed that under the RJA condition, the number of transitions from face to the target and the non-target objects were not significantly different between groups. With similar tools, Swanson et al. (2013) measured the eye movements of ASD (7.3±1.5) responding to Posner paradigm. Results showed no significant group difference on total fixation duration between the ASD and controls matched at receptive language ability, but significant group difference emerged after using first fixation duration as a measurement of gaze microstructure. Bedford et al. (2012) used the video clips depicting one actress gazed at one out of two targets to clarify the precursors of joint attention in infant at risk for the ASD at 7 and 13 months. They did not find any significant difference between groups in both visits, but individuals who got a diagnosis of ASD in 3-year old showed reduced looking time on the target object. Guillon et al. (2014) suggested that there might not be global joint attention deficits in the ASD population, but some more subtle difficulties. Thus, it is still an open question whether and how children with ASD show different joint attention skills from TD children.

In the previous studies, to define the fixation time on each area of interest (AOI) and the gaze shifts between different AOIs were the classic data analysis method. However, they only reflected the static aspect of the attention distribution and had difficulties to capture those dynamic features. Recently, researchers started to consider the complexity of transition distributions between AOIs such as Shannon's entropy (Krejtz et al. 2015) that reflects the degree of disorder. The more ordered the system is, the lower the value of Shannon's entropy is, and vice versa. Researchers introduced Shannon's entropy to analyze the complexity of the transition distribution between AOIs and the complexity of the resulting stationary distribution. It has shown a positive relationship with randomness of the gaze patterns between the AOIs. The exploration in this area could surely benefit the eye-tracking researches. In the joint attention experiments, the measured consecutive gaze points were expected to be logically related. Cristino et al. (2010) used a method called "ScanMatch", which was based on the longest common subsequence (LCS) algorithm, to find the largest common character sequence between the two given sequences of fixation generated by eye movements. They verified the feasibility of the algorithm by quantifying the similarity of visual fixation sequences while looking at familiar objects. In this paper, we introduce a new algorithm based on LCS that is capable of measuring the fully dynamic process of joint attention. The proposed algorithm is to quantify how the participant's gaze dynamically follow the given logic of the videos. The details would be elaborated in the "Method" section.

In the present study, we intended to use eye tracking to investigate whether the humanoid was able to improve the joint attention ability of the children with ASD from 3 to 6 years old. By utilizing a commercial humanoid NAO, this study compared the difference of joint attention behaviors induced by both the human and robot stimuli. Beyond the analysis of fixation time, gaze transitions, we induced a new LCS approach to analysis the dynamic eye-movement logic. The present study could reveal both the static and dynamic process underlying the joint attention behavior and compare the difference between the ASD and TD children. To the best of our knowledge, this is the first study using eye tracking to assess the gaze pattern induced by a robot and systematically comparing the effects of human and robot initiators on the gaze following. Based on the findings of the previous studies, we proposed the following hypothesizes for this study: (1) The ASD might spend less time on the agents (human or robot) face and the target than the TD group; (2) The robot could facilitate joint attention behavior similar as the human; (3) The ASD might make less gaze transition between the face and the target than the TD group; (4) The ASD might not follow the predefined logic of the videos as the TD group did.

## Method

### Participants

21 ASD and 22 TD children participated in this study. The children with ASD were recruited from an ASD special education institution. All the children with ASD had a formal diagnosis in certificated hospitals by professional pediatricians or psychiatrists based on *Diagnostic and Statistical Manual of Mental Disorders* 4$^{th}$ edition, text revision (American Psychiatric Association, 2000). All children with TD were recruited from a local kindergarten, with no ASD or other developmental disorders histories. Both groups came from two suburb districts nearby in Guangzhou. 3 children with ASD and 5 with TD could not pass the experiment due to the participants' reluctance to attend the experiment or technical reasons. The participants, who contributed more than two trails with less than 20% of total fixation time on the screen, were excluded for data analysis. Thus, the data of 3 ASD and 2 TD could not meet the criterion above, leading to 15 ASD (13 males and 2 females) and 15 TD (12 males and 3 females) final valid participants. The total fixation time percentage on the screen was similar between two groups and conditions.

Two groups matched at the chronological age and gender ratio (see Table 1). All of the participants had normal visual and hearing ability, and none of them had other diseases or took medicines which could influence the experiment results. The receptive language ability of the participants was assessed by the Chinese version of Peabody Picture Vocabulary Test (PPVT) (Biao and Xiaochun, 1990). The TD group had significantly higher PPVT scores than the ASD group, so the PPVT score would be treated as a covariance in further analysis. All the parents of participants provided informed consent before the experiment. The study protocol was approved by the Ethic Committee of University.

Table.1 demographic details of the final sample

|  | ASD | TD |  | p. |
|---|---|---|---|---|
| Age (in years) | 4.96±1.10 | 4.53±0.90 | $t = 1.16$ | 0.257 |
| PPVT | 97.65±24.64 | 119.40±14.42 | $t = 2.95$ | 0.006* |
| Male: female ratio | 13:2 | 12:3 | $\chi^2 = 0.24$ | 0.624 |
| Total fixation time percentage | 43.57%±20.22% | 47.38%±14.22% | $t = -0.60$ | 0.556 |

*: $p < 0.05$; PPVT: Peabody Picture Vocabulary Test

**Stimuli**

The stimuli consisted of two conditions: 1) "Human" condition that displayed a man sitting behind a table with a neutral facial expression, tried to induce the joint attention by turning his head towards one of the three truck toys in front of him; and 2) "Robot" condition displayed a social robot in the same circumstance as the human condition. Three toys were on the left, middle and right side of the table. In addition, there was a frame in the background considered as a distractor.

Each condition had four video clips and each one contained two segments: 1) In the greeting segment, the agent raised the head, looked directly at the participants and said "Hello, kid" in Chinese (2.5s); 2) In the joint attention segment, the agent started to turn the head to one of the three truck toys (i.e., the target object) and kept looking at the target (5~6s). The sequence of the gazing the target was left-middle-right-left for all the videos. Screen shots of video clip of each condition were shown on Fig.1 (A) & (B).

The robot used in this project was NAO. NAO is a programmable humanoid robot developed by Aldebaran Robotics Company. The robot is 58 centimeters high, and has 25 degrees of freedom. After careful programming, it could accomplish the segments described above.

**Procedure**

Each participant sat in a child-sized chair in front of the monitor in a quiet classroom, which all participants were familiar with. Before the formal experiment, participant got the chance to play and interact with robot for 10 minutes. If the ASD participants could not settle down, their parents were instructed to sit behind. The experimenter firstly showed a cartoon movie to attract their attention, after the participant had watched the cartoon clips quietly for more than one minute, a 5-point calibration procedure was conducted. An accurate calibration required participants to fixate within 1° of each fixation point. Then the formal experiment started, the videos were displayed on a 22" color LCD monitor at a distance of approximately 60 cm and subtended a visual angle of approximately 32° horizontally and 24° vertically. The screen resolution was 1920 × 1080 pixels. Then the participant was instructed to view the video clips freely. Half of the participants started with the human condition, and the other half started with the robot condition. The participant was always allowed to freely move their head position during the experiment but asked to "sit quite still". If the participant could not pass the calibration or the experiment session, the same experimental procedure would be repeated in the same setting at least two weeks later.

The eye movement was recorded by a remote screen-based Tobii X3-120 eye-tracker system (Tobii, Sweden). The frequency of recording is 120 Hz and the accuracy is 1° of visual angle. The accuracy of recording was maintained throughout the experiment as long as the participants kept their eyes within a virtual space measuring 20×20×20 cm. Moving outside the virtual space could cause recording a temporary stop, and returning the head to the correct position could re-start recording.

## Data analysis
### Definition of AOIs
Only the data during the joint attention stage were included in the analysis. By using the AOI tool in the Tobii Studio, we have defined 7 AOIs, namely face, body, three toys, frame and background (see Fig.1). Gaze data was extracted for each AOI.

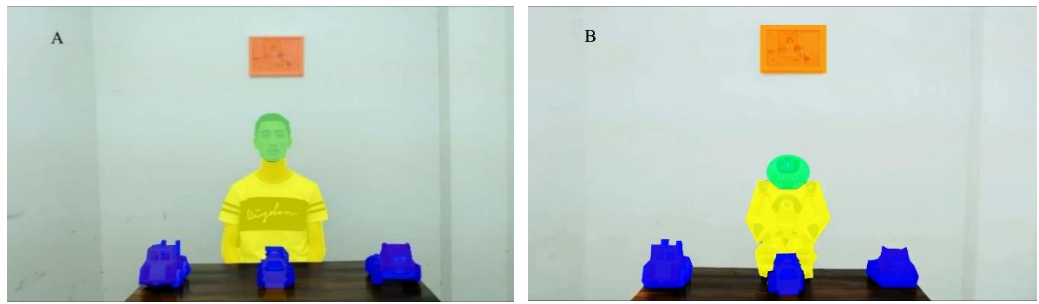

Fig.1 The display of stimuli and AOIs: (A) The "Human" condition, and (B) the "Robot" condition. The green area is the agent's face. The yellow area is the agent's body. The blue areas are three objects. The orange area is the frame. The rest of the screen is the background.

### Fixation analysis
A fixation was defined as eye-tracking points, which were within 1.5° of visual angle for 100ms or longer. The first fixation time was defined as the latency of the first fixation on each AOIs, and the fixation duration percentage was defined as the proportion of the time spent on each AOI over the total fixation time on the screen. The toy that gazed by the agent was defined as the target, and the other two toys were defined as the non-target. The fixation duration percentages of the target and the non-target were averaged across the whole experiment. Six AOIs would be included in the analysis: face, body, target, non-target, frame and background.

### Gaze Transitions
The number of gaze transitions was extracted between the face and the three toys with a homemade script written in Matlab and Java. The congruent gaze shifts were defined as the transitions between the face and the target (Fig. 2) and the incongruent gaze shifts were between the face and the non-target. The analysis focused on the number of congruent and incongruent gaze shifts directly.

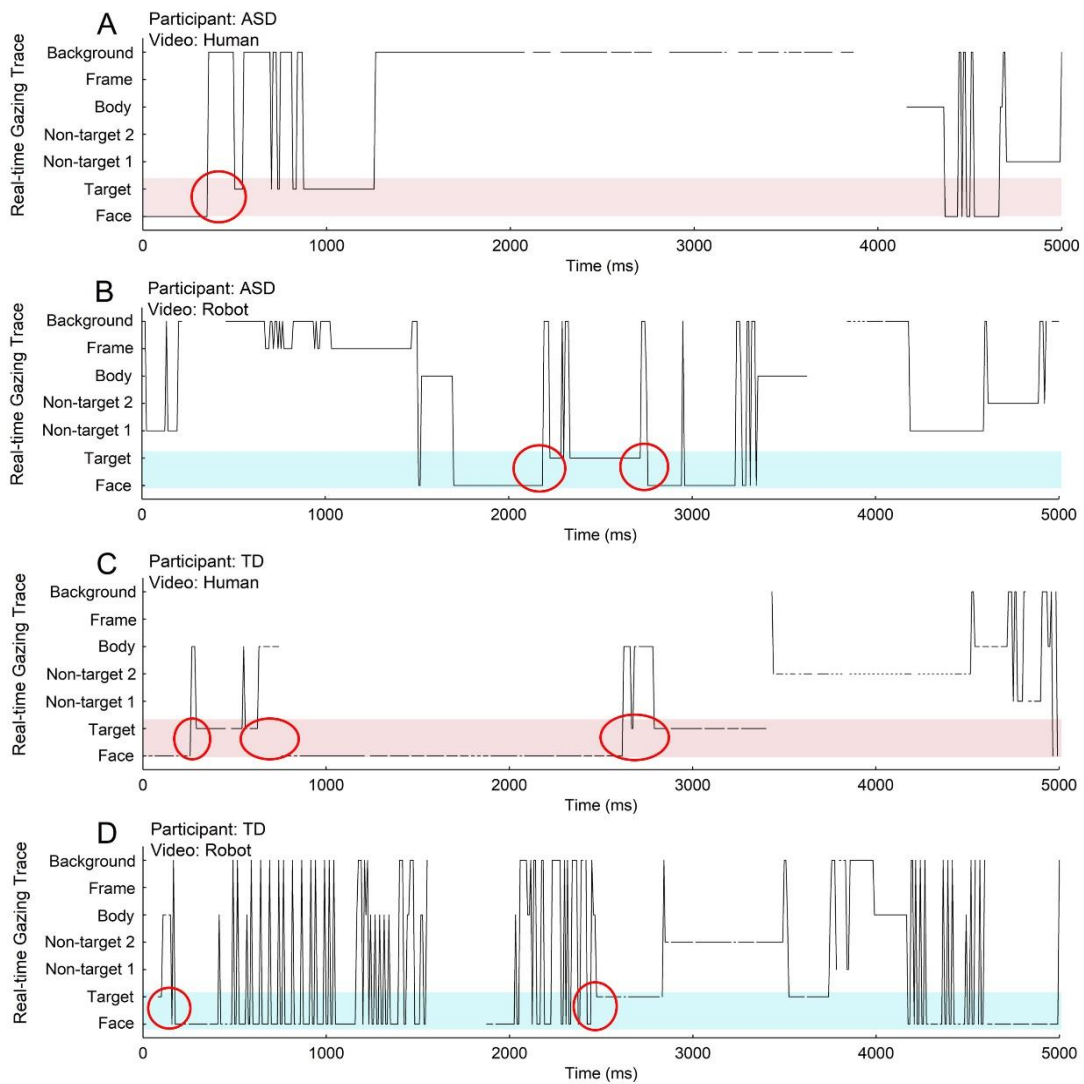

Fig.2 The real-time gazing traces of two participants (one child with ASD and one TD) during joint attention segment of the video. The Y axis lists different AOIs, and the X axis represents the time points. The black lines represent eye movement traces, where each point corresponds to one AOI measured at a certain time moment. A is for an ASD child in "Human" condition and B is for the same ASD child in "Robot" condition. C and D depict a TD child in "Human" and "Robot" conditions, respectively. The red-line circles point out the gaze transitions that happened between the agent's face and the target. The light red and light blue blocks cover the face and target regions, which are provided as the reference settings in LCS analysis, and the joint attention behavior, can be discovered in these areas.

## LCS analysis

In the present study, we introduced a longest common subsequence (LCS) algorithm for the data analysis of eye movements. In the present study, the considered joint attention contained: 1) the attention to the face of the agent, following the gaze direction of the agent and transiting the attention to the target, 2) the attention to the target, and maybe turning back to the face of the agent. The children who show one or more these aspects were considered to demonstrate joint attention skills. Therefore, the introduced LCS analysis not only focused on altering gaze between the face and the target but also the gaze purely at the face or the target, which were strongly related to the possible aspects of joint attention.

The LCS algorithm has been widely used in computational biology and human genome planning. In general, for two sequences $X = (X_1, X_2, ..., X_n)$ and $L = (L_1, L_2, ..., L_i)$, if $L = (X_{j_1}, X_{j_2}, ...X_{j_i})$, where $1 \leq J_1 < J_2 < ... < J_i < ... \leq J_n$, then $L$ is the subsequence of $X$. For any two sequences $X$ and $Y$, if $L$ is the subsequence of $X$ and $Y$, and if there is no other common subsequence of $X$ and $Y$ longer than $L$, then $L$ is the LCS of $X$ and $Y$. The LCS of these two given strings is not necessarily unique. Nevertheless, the length of the LCS is unique.

In the present study, the gaze point data (with 120Hz time resolution), which contained both spatial and temporal information, was extracted. We assigned each AOI to a certain value: "1" for "face", "2" for "body", "3" for "target", "4" for "non-target", "5" for "frame", and "6" for "background". Then we transformed the measured gaze trace into a sequence $X$. For example, if a participant's gaze trace measured during 48ms was as follow: Face/Face/Body/Target/Target/Target, then the transformed sequence would be 112333.

During the joint attention stage of the eye-tracking measurement, it was assumed that the participant should focus on either the agent's face or the target. The reference sequence defined in the present study reflected the intrinsic logic of the research stimuli of joint attention, while participant should focus on the agent's face, follow the gaze direction, and look at the target, possibly back and forth. Correspondently, we formed a reference sequence Y, which contained either "1" or "3", representing the agent's face or the target. In this case, each element in the sequence $Y$ was a set containing two values. We define $N$ as the length of $Y$, which was the same as $X$, namely both $X_i$ and $Y_i$ corresponded to the same time point of eye movements. The LCS algorithm applied to the present study was based on a recursive function. A two-dimensional matrix $M[i,j]$ could be derived according to Formula (1). The length of the LCS of X and Y equals to the element of M that has the largest value. The percentage presented in the later section for LCS score was the ratio of the length of the LCS of X and Y over the length of X (or Y).

$$M[i,j] = \begin{cases} 0, \text{if } i = 0 \text{ or } j = 0 \\ M[i-1, j-1] + 1, \text{if } i, j > 0 \text{ and } X_i \subseteq Y_j \\ \text{Max}\{M[i-1, j], M[i, j-1]\}, \text{if } i, j > 0 \text{ and } X_i \nsubseteq Y_j \end{cases} \quad (1)$$

## Results

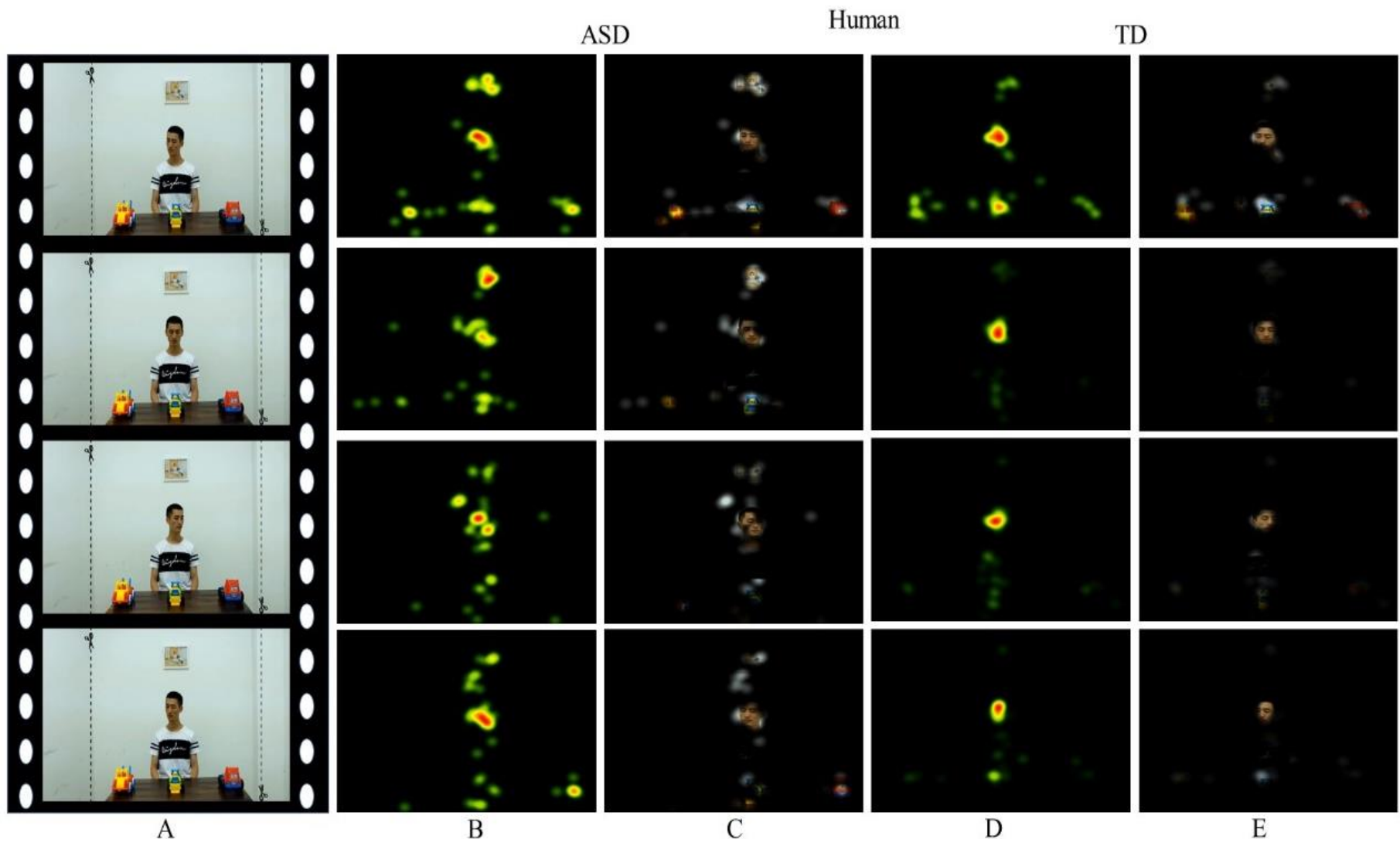

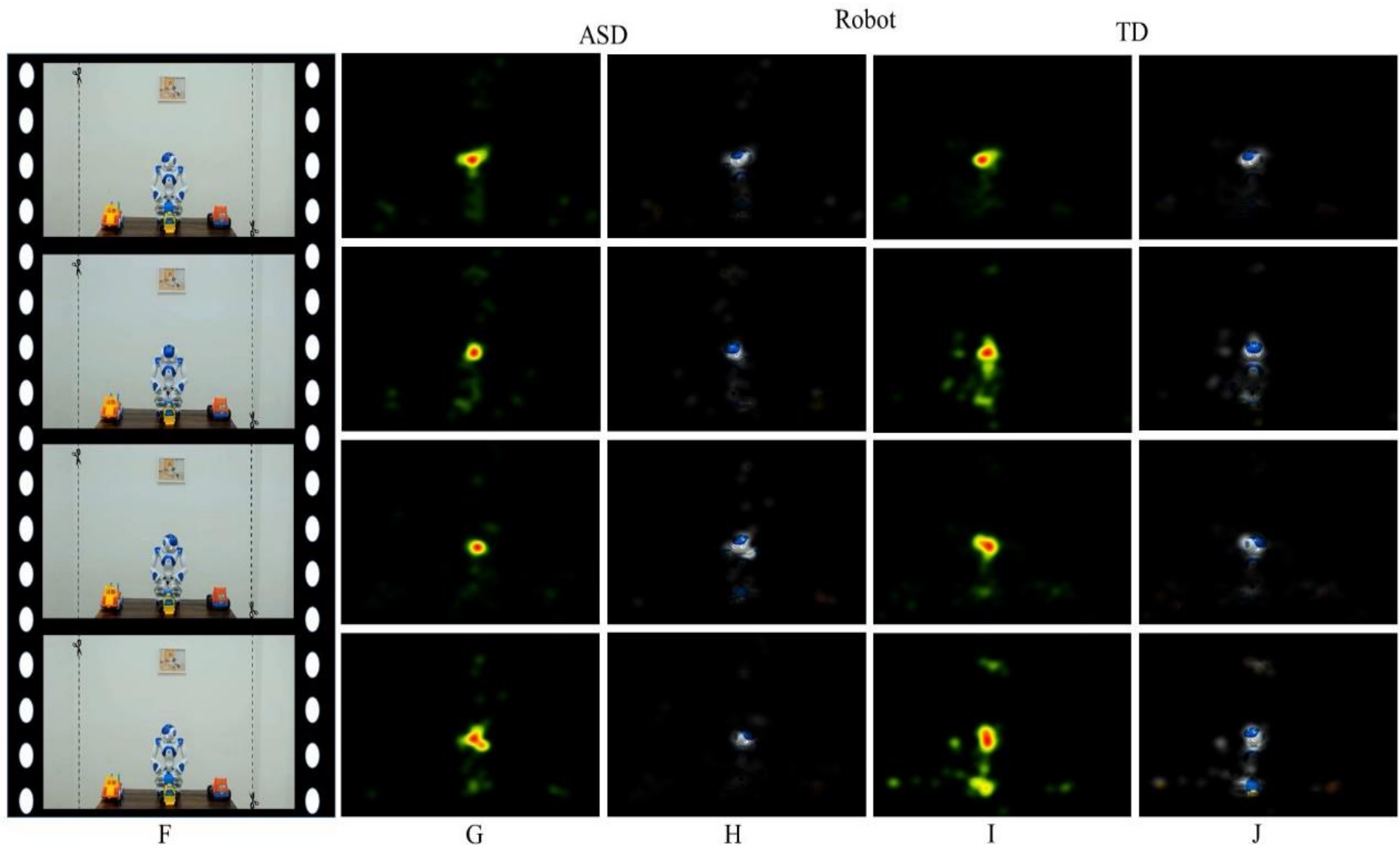

Fig.3 The visual fixation data during the joint attention stage. A and F depict a static frame from the 4 episodes sequentially from the joint attention stages. B and G are the averaged heat maps all the ASD participants in "Human" and "Robot" condition; D and I are the averaged heat maps for all the TD participants. Hotter colors denote greater density. C and H are the averaged heat maps scaled from black to transparent of the ASD participants in "Human" and "Robot" condition, overlaid on the original frame; E and J are the averaged heat maps scaled from black to transparent of the TD participants.

Figure 3 depicts the visual fixation data during 4 joint attention stages, from which we could derive some preliminary findings. In the "Human" condition, the TD participants were more likely to look at the agent's face and the target. On the contrary, the ASD participants spent their time on the agent's face, body, three objects and the frame. Meanwhile, in the "Robot" condition, both groups spent more time on the robot's face and body, and less on the target. These findings would undergo tests in the following analysis.

For all the dependent variables mentioned above, a Repeated Measures ANOVA was conducted to determine the effect of independent variable, group (ASD vs. TD) was the between group variable, stimuli type ("Human" vs. "Robot") and AOI (face, body, target, non-target, frame and outside) were the within group variables. *Post-hoc* analysis was conducted if a main effect or an interaction effect was significant. The statistic process was performed using SPSS 20.0.

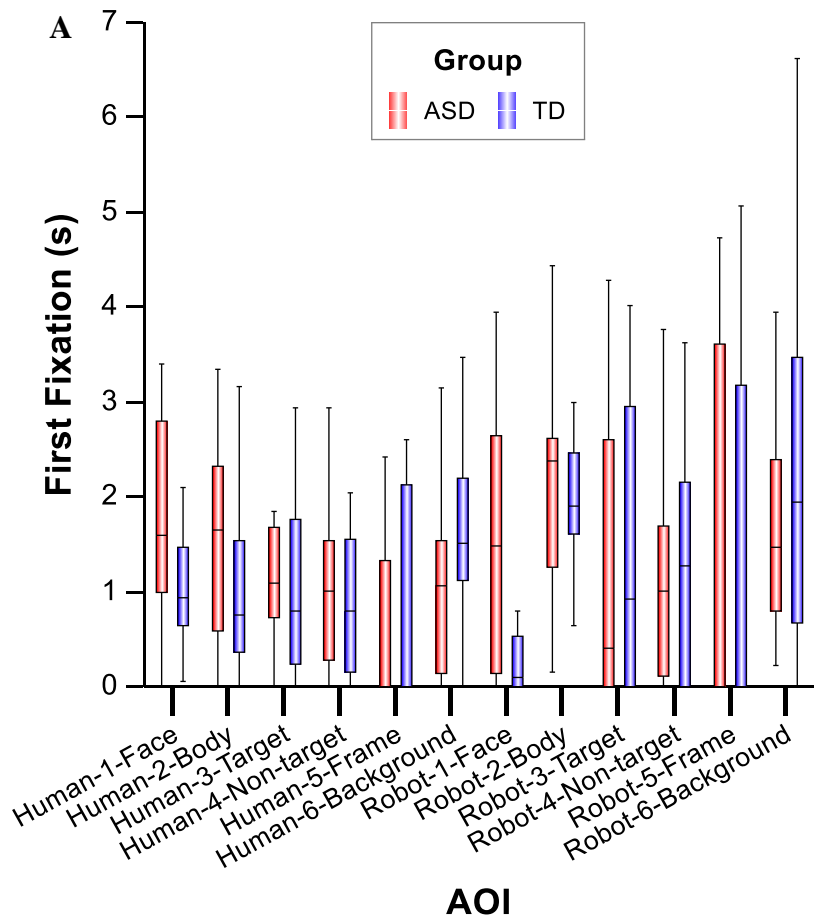

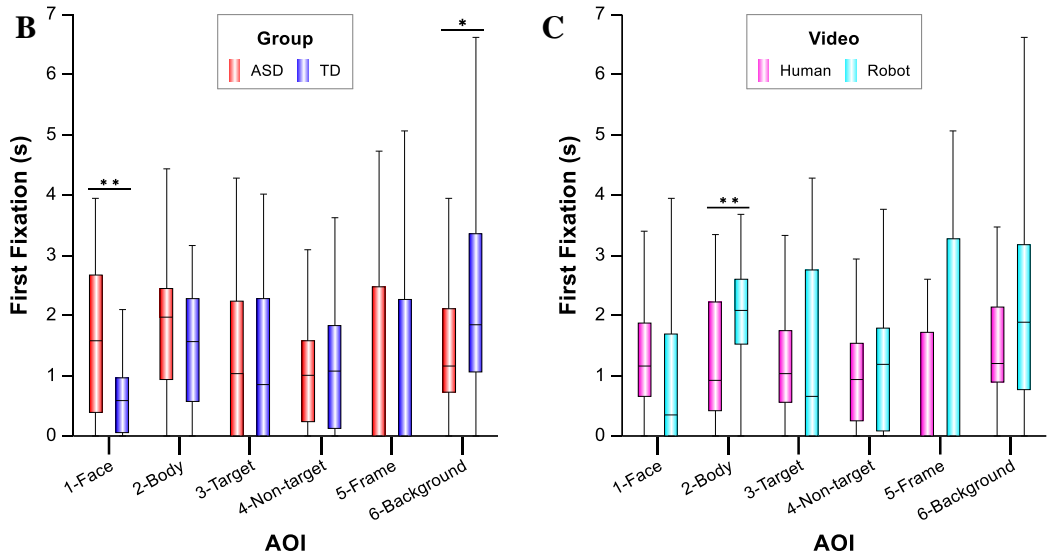

Fig. 4 (A) The first fixation time (Seconds) on different AOIs of the ASD and TD participants. (B) The results of the *post-hoc* analysis of the interaction effect of group and AOI, which indicates that the TD group looked at face faster than ASD group, while ASD looked at background faster than TD group. (C) The results of the interaction effect between stimuli type and AOI, which demonstrates that all participants looked at human body faster than robot body.

**First fixation time**

To determine if both groups were equally likely to be looking at the face before the gaze shift occurred, the Repeated Measures ANOVA of the first fixation time of each AOIs was conducted. The main effect stimuli type ($F(1, 28) = 5.502$, $p = 0.026$, partial $\eta^2 = 0.164$), and AOI ($F(1, 28) = 2.989$, $p = 0.014$, partial $\eta^2 = 0.096$), and the interaction effect between group and AOI ($F(5, 28) = 3.164$, $p = 0.010$, partial $\eta^2 = 0.102$), between stimuli type and AOI ($F(1, 28) = 2.769$, $p = 0.020$, partial $\eta^2 = 0.090$) were significant (Fig.4 (A)). The *post-hoc* analysis revealed that participants gazed at the human agent faster than the robot ($p = 0.026$), and the TD group looked at the agent's face faster than the ASD group ($p = 0.001$), but spent more time to fixate on the background ($p = 0.049$) (Fig.4 (B)). Moreover, the participants looked at the human body faster than the robot body ($p = 0.008$) (Fig.4 (C)). All the effect remained significant after considering PPVT scores as the covariance. No other main effects or interaction effects reached significance.

**Percentage of fixation time**

The repeated ANOVA revealed a significant main effect of AOI ($F (5, 28) = 44.00$, $p<10^{-7}$, partial $\eta^2 = 1.000$), a significant interaction effect between AOI and group ($F (5, 28) = 5.63$, $p =10^{-3}$, partial $\eta^2 = 0.946$), and between AOI and stimuli type ($F (5, 28) = 3.08$, $p = 0.03$, partial $\eta^2 = 0.684$). The main effect of group, stimuli type, the interaction effect between group and video type and the three level interaction effect did not reach statistical significance. P values and power values were adjusted with Huynh Feldt Epsilon method. The *post-hoc* analysis (with a Bonferroni correction) showed that the TD group fixated longer on the agent's face ($p = 0.002$) than the ASD group, but the ASD group gazed on the frame ($p = 0.045$) and non-target ($p = 0.036$) longer than the TD group (Fig.5 (B)). All participants spent more time on the face area in the "Robot" condition ($p = 0.038$), and more on the target ($p = 0.004$) and non-target in the "Human" condition ($p = 0.045$) (Fig.5 (C)).

After considering the covariance, only the significant interaction effect between AOI and group remained ($F(5, 28) = 5.02$, $p = 0.02$, partial $\eta^2 = 0.157$).

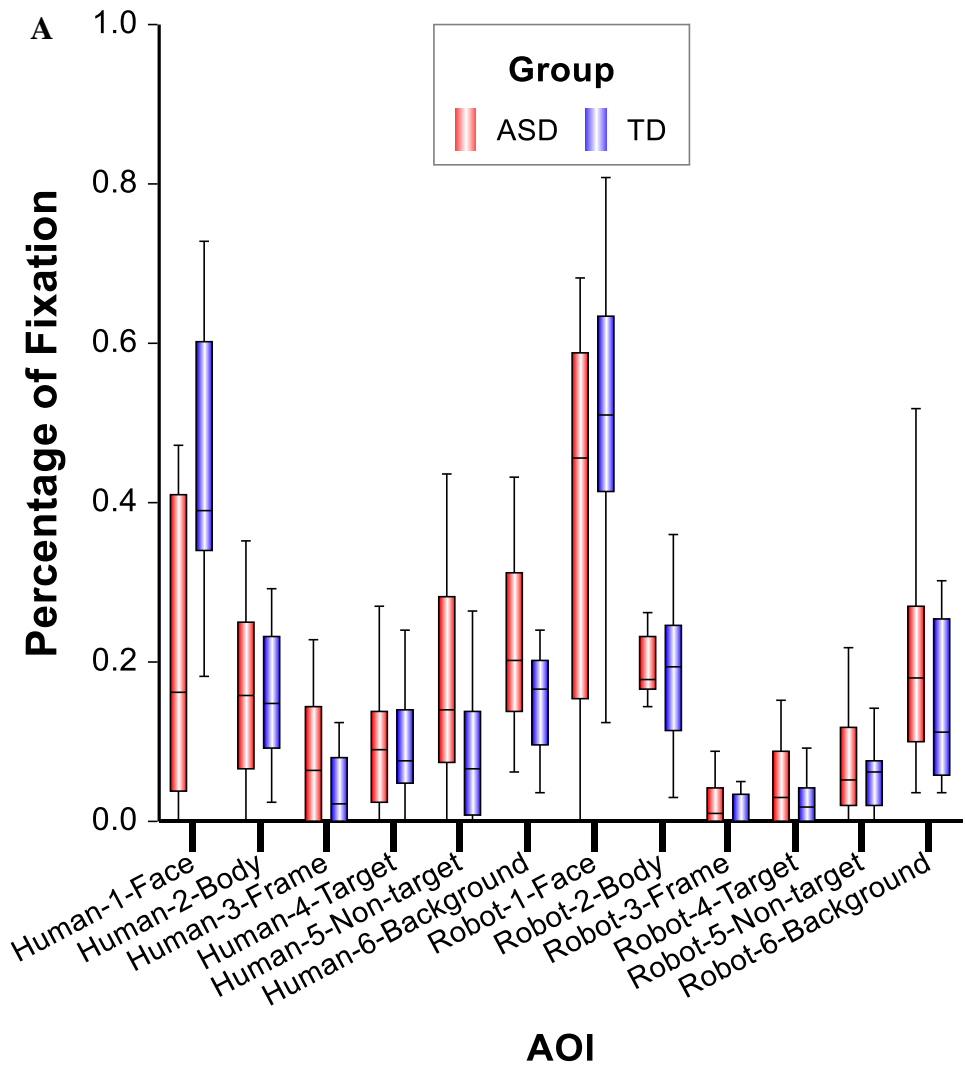
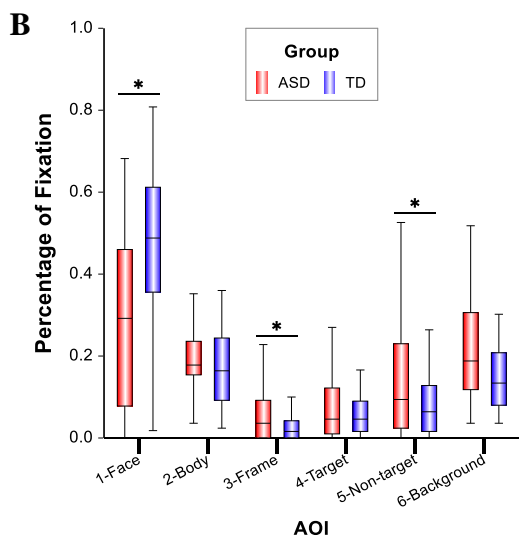
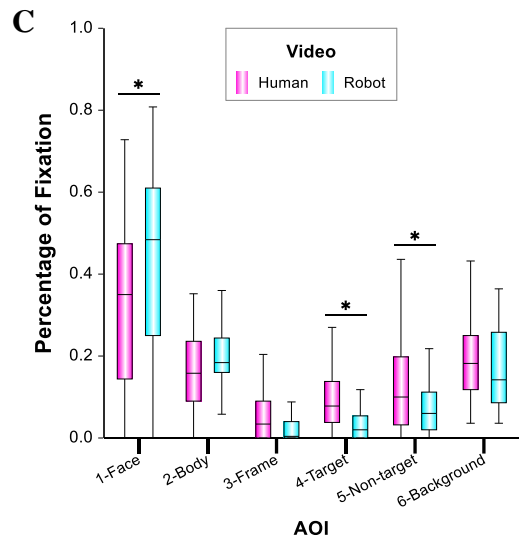

Fig. 5 (A) The percentage of fixation time on different AOIs of the ASD and TD participants. (B) The results of the *post-hoc* analysis of the interaction effect of group and AOI, which indicates that the TD group fixated longer on the agent face than the ASD group, but the ASD group looks on the frame and non-target longer than the TD group. (C) The results of the interaction effect between stimuli type and AOI, which demonstrates that all participants spend more time on face area in the "Robot" condition ($p = 0.038$), and more on target and non-target in the "Human" condition.

## Gaze transition

A 3×2 ANOVA (group, stimuli type, target type) did find some significant difference between two groups ($F (1, 28) = 9.282$, $p = 0.005$, partial $\eta^2 = 0.154$), but no significant effect between stimuli types ($F (1, 28) = 1.098$, $p = 0.304$, partial $\eta^2 = 0.038$), or target type ($F (1, 28) = 1.844$, $p = 0.185$, partial $\eta^2 = 0.062$) on the gaze transition number. Meanwhile, there was no interaction effect between group and stimuli types ($F(1, 28) = 0.962$, $p = 0.335$, partial $\eta^2 = 0.033$), group and target type ($F (1, 28) = 0.014$, $p = 0.908$, partial $\eta^2 < 0.001$), stimuli type and target type ($F(1, 28) = 1.196$, $p = 0.283$, partial $\eta^2 = 0.041$), or group\ stimuli type and target type ($F(1, 28) = 0.159$, $p = 0.699$, partial $\eta^2 = 0.007$). The TD children showed a significant higher congruent gaze transition number than the children with ASD. This main effect still survived after controlling the influence of PPVT scores ($F (1, 28) = 8.335$, $p = 0.008$, partial $\eta^2 = 0.236$).

## LCS

The repeated ANOVA result revealed a significant main effect of group ($F (1, 28) = 11.18$, $p = 0.002$, partial $\eta^2 = 0.898$). The *post-hoc* analysis revealed that the TD children have significantly higher LCS scores than the children with ASD, which implies the TD could better follow the logic of the video. Even using PPVT as a covariance, this effect was still significant ($F (1, 28) = 8.710$, $p = 0.006$, partial $\eta^2 = 0.244$). The main effect of stimuli types ($F (1, 28) = 1.852$, $p = 0.184$, partial $\eta^2 = 0.062$), interaction effects between group and video ($F (5, 28) = 1.85$, $p = 0.184$, partial $\eta^2 = 0.259$), and between stimuli types and groups ($F (1, 28) = 0.913$, $p = 0.347$, partial $\eta^2 = 0.032$) was not statistically significant.

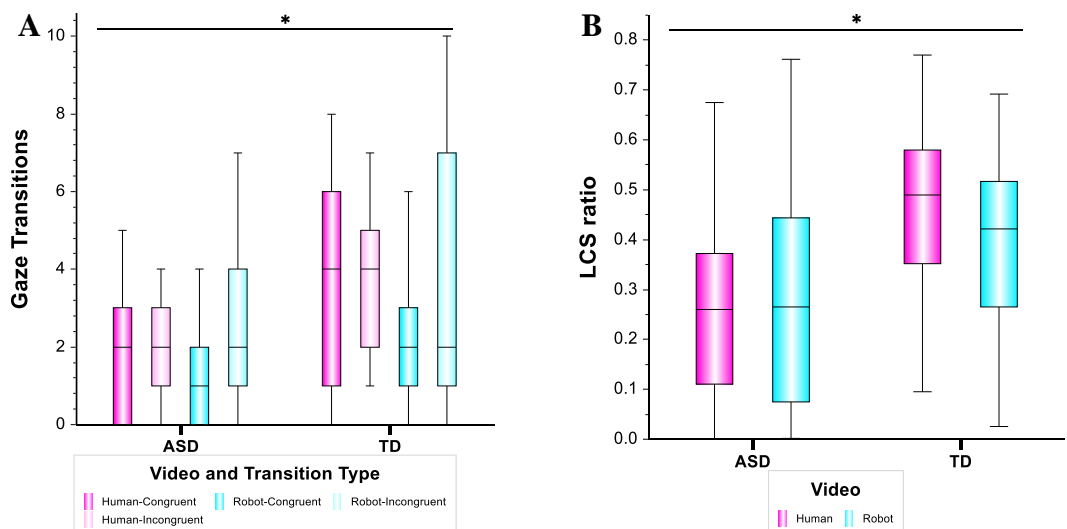

Fig. 6 The number of gaze transition and LCS scores of the children with ASD and TD children. (A) Gaze transition between different groups and stimuli types, (B) LCS scores between different groups and stimuli types.

## Discussions

**Joint attention deficits of children with ASD**

Given robot intervention was a promising tool for the ASD population, little was known about how the children with ASD responded to robot's gaze direction. The present study combined a commercial social robot with eye-tracking technology to provide evidence about the effect of robot on the children with ASD. Firstly, the present research confirmed that ASD did show joint attention deficits. The ASD participants made less gaze transitions between the initiator and the targets than the TD group. The reason may be that the ASD had difficulties in capturing and utilizing the social meaning of face, leading to abnormal attention distribution towards face (for both robot and human conditions). This could be confirmed by the fact that the ASD group fixated slower and less on the agent's face, and faster and longer on the frame and non-target than the TD group. Previous studies have indicated that ASD population showed abnormal face scanning patterns, especially less fixations on eye regions, and this visual pattern affected the extraction of social information from face (Chawarska and Shic, 2009). The joint attention behaviors required process of social signal (gaze direction) of initiator and connected it with the target, and the current study demonstrated this difficulty of the ASD participants in a behavioral level. Zhu et al. (2015) used functional near-infrared spectroscopy to measure the neural activity of the children with ASD (8.75 ± 1.34) to a video depicting joint and non-joint attention task. Results exhibited that the ASD showed reduced activation in prefrontal cortex and abnormal asymmetry interhemispheric connectivity in the joint attention condition compared to the TD controls, which provided insights on the functional brain activities underling the difference of joint attention behaviors.

**LCS analysis and its implications**

The LCS result has demonstrated that the ASD could not follow the given logic of the stimuli as the TD did, which was an important complementary to the fixation and gaze transition results. The analysis of eye tracking data based on the fixation time and gaze transition is not capable of capturing the dynamic feature of the joint attention task. The LCS algorithm, however, is a synthetic way to reveal the characteristic of the entire eye movement, covering fixation and gaze transition. It did not focus on the mean value of fixation time or numbers of gaze transition, but the dynamic process of joint attention. By means of the LCS algorithm, the similarity between the eye movement sequences of participants and the reference sequence could be calculated by traversing the two sequences. Joint attention behavior was underlining the reference sequence transition between the head and the target region, and the algorithm could calculate the similarity between the reference sequence and eye movement sequence of different groups. As a consequence, the LCS result could clarify the deficit of joint attention among children with ASD based on the logic of stimuli. Shic et al. (2008) also used a similar entropy-based methodology to measure the eye movement of the ASD to static picture, and found some new results that beyond the traditional fixation analysis, which surely indicated that it could be a promising tool in the future.

In the present study, the result of fixation analysis and the LCS analysis was slightly different, which may derive from two reasons. Basically, one fixation was the sum of several continuous gaze points, and the fixation time in particular AOI was the sum of fixations in that AOI. This was a continuous process of dimension reduction. The LCS, however, used the gaze points directly, which contained the spatial and temporal information of eye movement, and was assumed to better describe both the static and the dynamic feature of eye movements. Another reason maybe in the present study, the reference sequence predefined was relatively simple, leading to that the LCS scores are equivalent to the sum of the gazing on agent's face and targets. Because the exploration of LCS algorithm which applied to joint attention was just started, this logic was still kind of simple and may not be the perfect approximation of joint attention. However, the LCS method itself was universal and could be applied under much more complex situations. The reference sequence could be modified accordingly when the logic of joint attention stimuli was changed. Furthermore, the LCS could be applied to the case with more complicated logics for other cognitive process, where a dynamic process was required to be measured and quantified.

**The effect of robot on joint attention**
The present study found that the participants showed more interests on the robot's face, but spent more time on the target and non-target in the "Human" condition. This could be interpreted in two aspects. Firstly, the robot could attract the attention of the participants successfully. The result mentioned above has demonstrated that the ASD tended to avoid or ignore the human's face, which could somewhat account for the abnormal social attention of the ASD. It was important to ensure the children with ASD focused on the face and its social signals during diagnosis and intervention, but it was always a difficult task. The Robot, however, could benefit from its unique advantage and may play a particular role in the ASD diagnosis and intervention in the future. Damm et al. (2013) has found that the ASD performed more eye contacts and fixations in the robot face than the human face. Neuroimaging study also reported similar brain activation patterns when processing robot face between the ASD and TD groups (Jung et al. 2016). Secondly, the robot seemed to affect the attention distribution towards the target. Some previous researches did find that the ASD turned their attention to the robot instead of the target object, which seemed to imply that the robot might become the distractor in the joint attention process. Nevertheless, our results also showed that the participants performed similar in gaze transition and LCS scores in the "Robot" condition and "Human" condition. It was plausible to assume that robot might show a negative effect on the fixation time on the target, but the robot still could prompt gaze transition like the human agent, and the ASD participants could equally understand the social logic in the "Robot" condition. This fact drew a contemporary conclusion that joint attention behavior induced by the robot and human may show some differences in details, but few evidence could support an essential difference between the effect of robot and human agents on joint attention.

Although there were just some preliminary results addressing the clinic use of the robot in ASD intervention and it was much too early to conclude the final role of robot intervention, our results could support the feasibility of robot in ASD intervention for joint attention. The present study found that ASD could understand the joint attention logic induced by the robot. Furthermore, the previous research has also proved that the ASD activated similar cerebral areas when interacting with the robot compared to the TD controls interacting with the intentional agents (Chaminade et al., 2012). These outcomes implied that the robot might be qualified to be a social partner for children with ASD. With carefully environmental settings and process arrangements, interactions between the robot and ASD could clarify the social logic underneath, and extend to real life context (Pennisi et al., 2016). Without doubt, it did not mean to replace the human with the robot. Questions remained, though, what were the limitations when applying the robot in longitudinal intervention, but the robot could be emerged into the present ASD intervention system, and take roles like social mediators between the ASD and therapists, which would be a great relief to the country that are lack of professionals (Coeckelbergh el at. 2016).

**Limitations**

There were some limitations. First, despite of its humanoid appearance, the robot NAO still could not fully convey the social and emotional cues like human. To eliminate the influence of such factors, our experimental design tried to use minimum language, and asked the actor to keep his face pale for the whole process. This could affect the ecological validity of the stimuli, and should be modified by on-site experiment in the future. Secondly, the LCS algorithm still had potential to better describe the joint attention behavior. The current LCS algorithm did not take the gaze latency into consideration, and moreover, this algorithm still could not separate the detailed difference between conditions. This required different weight on different eye movement patterns. Finally, the present research only focused on the performance of ASD in one experimental session. It could provide limited information about the reaction of the children with ASD in a complete intervention course. This was also the insufficiency of many other researches. It called for clarification on how to integrate the robot intervention with generally accepted principles, and form a integrative and systematic intervention process.

# Reference

Anzalone, S. M., Tilmont, E., Boucenna, S., Xavier, J., Jouen, A.-L., Bodeau, N., et al. (2014). How children with autism spectrum disorder behave and explore the 4-dimensional (spatial 3D+ time) environment during a joint attention induction task with a robot. *Research in Autism Spectrum Disorders, 8*(7), 814-826.

American Psychiatric Association. (2000). *Diagnostic and Statistical Manual of Mental Disorders, Fourth Edition, Text Revision (DSM-IV-TR)*. Washington DC: American Psychiatric Press.

American Psychiatric Association. (2013). *Diagnostic and Statistical Manual of Mental Disorders, 5th Edition*. Washington DC: American Psychiatric Press.

Bedford, R., Elsabbagh, M., Gliga, T., Pickles, A., Senju, A., Charman, T., & Johnson, M. H. (2012). Precursors to social and communication difficulties in infants at-risk for autism: Gaze following and attentional engagement. *Journal of autism and developmental disorders, 42*(10), 2208-2218.

Bekele, E., Crittendon, J. A., Swanson, A., Sarkar, N., & Warren, Z. E. (2014). Pilot clinical application of an adaptive robotic system for young children with autism. *Autism, 18*(5), 598-608.

Biao, S., & Xiaochun, M. (1990). The Revision of Trail Norm of Peabody Picture Vocabulary Test Revised (PPVT-R) in Shanghai Proper. *Information on Psychological Science*, 22-27.

Billeci, L., Narzisi, A., Campatelli, G., Crifaci, G., Calderoni, S., Gagliano, A., et al. (2016).

Disentangling the initiation from the response in joint attention: an eye-tracking study in toddlers with autism spectrum disorders. *Translational psychiatry, 6*(5), e808.

Chaminade, T., Da Fonseca, D., Rosset, D., Lutcher, E., Cheng, G., & Deruelle, C. (2012, September). Fmri study of young adults with autism interacting with a humanoid robot. In *RO-MAN, 2012 IEEE* (pp. 380-385). IEEE.

Charman, T. (2003). Why is joint attention a pivotal skill in autism? *Philosophical Transactions of the Royal Society of London B: Biological Sciences, 358*(1430), 315-324.

Chawarska, K., & Shic, F. (2009). Looking but not seeing: Atypical visual scanning and recognition of faces in 2 and 4-year-old children with autism spectrum disorder. *Journal of Autism and Developmental Disorders, 39*(12), 1663-1672.

Centers for Disease Control and Prevention. (2016). Prevalence and Characteristics of Autism Spectrum Disorder Among Children Aged 8 Years - Autism and Developmental Disabilities Monitoring Network, 11 Sites, United States, 2012. *Morbidity and Mortality Weekly Report. Surveillance Summaries (Washington, D.C. : 2002), 65*(3), 1–23.

Coeckelbergh, M., Pop, C., Simut, R., Peca, A., Pintea, S., David, D., & Vanderborght, B. (2016). A survey of expectations about the role of robots in robot-assisted therapy for children with asd: Ethical acceptability, trust, sociability, appearance, and attachment. *Science and engineering ethics, 22*(1), 47-65.

Cook, J., Swapp, D., Pan, X., Bianchi-Berthouze, N., & Blakemore, S.-J. (2014). Atypical interference effect of action observation in autism spectrum conditions. *Psychological medicine, 44*(4), 731-740.

Cristino, F., Mathôt, S., Theeuwes, J., & Gilchrist, I. D. (2010). ScanMatch: A novel method for comparing fixation sequences. *Behavior research methods, 42*(3), 692-700.

Damm, O., Malchus, K., Jaecks, P., Krach, S., Paulus, F., Naber, M., et al. (2013). *Different gaze behavior in human-robot interaction in Asperger's syndrome: An eye-tracking study.* Paper presented at the RO-MAN, 2013 IEEE.

Diehl, J. J., Schmitt, L. M., Villano, M., & Crowell, C. R. (2012). The clinical use of robots for individuals with autism spectrum disorders: A critical review. *Research in Autism Spectrum Disorders, 6*(1), 249-262.

Falck-Ytter, T., Fernell, E., Hedvall, Å. L., von Hofsten, C., & Gillberg, C. (2012). Gaze performance in children with autism spectrum disorder when observing communicative actions. *Journal of autism and developmental disorders, 42*(10), 2236-2245.

Gredebäck, G., Fikke, L., & Melinder, A. (2010). The development of joint visual attention: a longitudinal study of gaze following during interactions with mothers and strangers. *Developmental science, 13*(6), 839-848.

Guillon, Q., Hadjikhani, N., Baduel, S., & Rogé, B. (2014). Visual social attention in autism spectrum disorder: Insights from eye tracking studies. *Neuroscience & Biobehavioral Reviews, 42*, 279-297.

Jung, C. E., Strother, L., Feil-Seifer, D. J., & Hutsler, J. J. (2016). Atypical asymmetry for processing human and robot faces in autism revealed by fNIRS. *PloS one, 11*(7), e0158804.

Kim, E. S., Berkovits, L. D., Bernier, E. P., Leyzberg, D., Shic, F., Paul, R., & Scassellati, B. (2013). Social robots as embedded reinforcers of social behavior in children with autism. *Journal of autism and developmental disorders, 43*(5), 1038-1049.

Krejtz, K., Duchowski, A., Szmidt, T., Krejtz, I., González Perilli, F., Pires, A., et al. (2015). Gaze

transition entropy. *ACM Transactions on Applied Perception (TAP), 13*(1), 4.

Mundy, P., & Gomes, A. (1998). Individual differences in joint attention skill development in the second year. *Infant behavior and development, 21*(3), 469-482.

Mundy, P., Sigman, M., Ungerer, J., & Sherman, T. (1986). Defining the social deficits of autism: The contribution of non-verbal communication measures. *Journal of child psychology and psychiatry, 27*(5), 657-669.

Mundy, P., Sullivan, L., & Mastergeorge, A. M. (2009). A parallel and distributed-processing model of joint attention, social cognition and autism. *Autism Research, 2*(1), 2-21.

Pennisi, P., Tonacci, A., Tartarisco, G., Billeci, L., Ruta, L., Gangemi, S., & Pioggia, G. (2016). Autism and social robotics: A systematic review. *Autism Research, 9*(2), 165-183.

Riby, D. M., Hancock, P. J., Jones, N., & Hanley, M. (2013). Spontaneous and cued gaze-following in autism and Williams syndrome. *Journal of neurodevelopmental disorders, 5*(1), 13.

Shamsuddin, S., Yussof, H., Mohamed, S., Hanapiah, F. A., & Ismail, L. I. (2013, November). Stereotyped behavior of autistic children with lower IQ level in HRI with a humanoid robot. In *Advanced Robotics and its Social Impacts (ARSO), 2013 IEEE Workshop on* (pp. 175-180). IEEE.

Shic, F., Chawarska, K., Bradshaw, J., & Scassellati, B. (2008, August). Autism, eye-tracking, entropy. In *Development and learning, 2008. ICDL 2008. 7th IEEE International Conference on* (pp. 73-78). Ieee..

Sun, X., Allison, C., Matthews, F. E., Zhang, Z., Auyeung, B., Baron-Cohen, S., & Brayne, C. (2015). Exploring the underdiagnosis and prevalence of autism spectrum conditions in Beijing. *Autism Research, 8*(3), 250-260.

Swanson, M. R., Serlin, G. C., & Siller, M. (2013). Broad autism phenotype in typically developing children predicts performance on an eye-tracking measure of joint attention. *Journal of autism and developmental disorders, 43*(3), 707-718.

Thorup, E., Nyström, P., Gredebäck, G., Bölte, S., & Falck-Ytter, T. (2016). Altered gaze following during live interaction in infants at risk for autism: an eye tracking study. *Molecular autism, 7*(1), 12.

Warren, Z. E., Zheng, Z., Swanson, A. R., Bekele, E., Zhang, L., Crittendon, J. A., et al. (2015). Can robotic interaction improve joint attention skills? *Journal of autism and developmental disorders, 45*(11), 3726-3734.

Yi, L., Fan, Y., Quinn, P.C., Feng, C., Huang, D., et al. (2013). Abnormality in face scanning by children with autism spectrum disorder is limited to the eye region: Evidence from multi-method analyses of eye tracking data. *Journal of Vision*, 13(10): 5-5.

Zhu, H., Li, J., Fan, Y., Li, X., Huang, D., & He, S. (2015). Atypical prefrontal cortical responses to joint/non-joint attention in children with autism spectrum disorder (ASD): A functional near-infrared spectroscopy study. *Biomedical optics express, 6*(3), 690-701.